\documentclass[aps,prd,preprint]{revtex4}

\usepackage{epsfig}

\begin{document}

\author{Yu Jiang$^{1}$, Hua Li$^{1}$, Shi-song Huang$^{2}$, Wei-min Sun$^{1,3}$ and Hong-shi Zong$^{1,3}$}
\address{$^{1}$ Department of Physics, Nanjing University, Nanjing 210093, China}
\address{$^{2}$ Department of Physics, Nanjing Normal University, Nanjing 210097, China}
\address{$^{3}$ Joint Center for Particle, Nuclear Physics and Cosmology, Nanjing 210093, China}
\title{The Equation of State and Quark Number Susceptibility in
Hard-Dense-Loop Approximation}
\begin{abstract}
Based on the method proposed in [ H. S. Zong, W. M. Sun, Phys. Rev.
\textbf{D 78}, 054001 (2008)], we calculate the equation of state
(EOS) of QCD at zero temperature and finite quark chemical potential
under the hard-dense-loop (HDL) approximation. A comparison between
the EOS under HDL approximation and the cold, perturbative EOS of
QCD proposed by Fraga, Pisarski and Schaffner-Bielich is made. It is
found that the pressure under HDL approximation is generally smaller
than the perturbative result. In addition, we also calculate the
quark number susceptibility (QNS) at finite temperature and finite
chemical potential under hard-thermal/dense-loop (HTL/HDL)
approximation and compare our results with the corresponding ones in the previous literature.
\bigskip


\bigskip

\noindent E-mail: zonghs@chenwang.nju.edu.cn.

\bigskip

\noindent PACS Number(s): 12.38 Mh, 11.10 Wx

\end{abstract}

\maketitle

The investigation of the equation of state (EOS) for cold and dense
strongly interacting matter and its consequence for the possible
phases of quantum chromodynamics (QCD) plays a crucial role in the
study of neutron stars in astrophysics \cite{N1,N2}. At present
lattice QCD provides the most powerful tool in studying QCD
thermodynamics. While the Monte Carlo simulation works well and has
achieved great success in dealing with finite temperature QCD
thermodynamics, it encounters the notorious sign problem in the
situation of finite quark chemical potential and cannot be applied
\cite{Halasz}. To circumvent the sign problem, several methods have
been proposed. Among them are the reweighting technique
\cite{Fodor}, analytic continuation from imaginary chemical
potential \cite{Forcrand, D'Elia} and Taylor expansion in $\mu$
around $\mu=0$ \cite{Allton1,Allton2,Sourendu}. Although these
methods have achieved some success in treating finite $\mu$ physics,
they are far from being complete and even suffer from some limitations and
drawbacks \cite{Stephanov0}. So we expect that continuum model
studies should be complementary to the lattice simulations in the
exploration of QCD thermodynamics at finite $\mu$. At high enough temperature ($T$)
and/or the quark chemical potential ($\mu$), the
hard-thermal/dense-loop (HTL/HDL) approximation is thought to be a
good approximation for QCD \cite{Braaten}. In the present paper, based on the method
proposed in Ref. \cite{Z1}, we try to calculate the EOS of QCD at zero temperature and high enough
quark chemical potential under HDL approximation. Since the HDL approximation is a good
approximation for QCD at high enough quark chemical potential, the
EOS derived with the aid of it should have a relatively good QCD
foundation and one expects that it can be applied to the study of
neutron star. In addition, with the method proposed in the present paper we can calculate another
important quantity for the study of QCD phase transition, the quark number
susceptibility (QNS), in the HTL/HDL approximation. The QNS measures the intrinsic statistical
fluctuations in a system close to thermal equilibrium and is thought
to play an important role in identifying the critical end point in
the QCD phase diagram \cite{Stephanov,Hatta1,Foder,Stephanov1,Lungwitz,Schaefer,He1,He2}. 

Let us start with the renormalized partition function of QCD at zero
$T$ and finite $\mu$ which reads
\begin{eqnarray}
{\cal{Z}}[\mu]&=&\int{\cal{D}}\bar{q_R}{\cal{D}}q_R{\cal{D}}
A_R~\exp\left\{-S_{R}[\bar{q}_R,q_R,A_R]+\int d^4x~\mu
Z_2\bar{q}_R(x)\gamma_{4}q_R(x)\right\},
\end{eqnarray}
where $S_{R}[\bar{q}_R,q_R,A_R]$ is the standard renormalized
Euclidean QCD action with $q_R$ being the renormalized quark fields
with three flavors and three colors, $Z_2=Z_2(\zeta^2,\Lambda^2)$ is
the quark wave-function renormalization constant ($\zeta$ is the
renormalization point and $\Lambda$ is the regularization
mass-scale). Here we leave the ghost field term and its integration
measure to be understood.

The pressure density ${\cal{P}}(\mu)$ is given by
\begin{equation}
{\cal{P}}(\mu)=\frac{1}{{\cal{V}}}~\ln\cal{Z}[\mu],
\end{equation}
where ${\cal{V}}$ is the four-volume normalizing factor. The above
expression for ${\cal{P}}(\mu)$ is just the EOS. From the pressure
density one immediately obtains the quark number density:
\begin{eqnarray}
\rho(\mu)=\frac{\partial\cal{P}(\mu)}{\partial\mu}.
\end{eqnarray}
 According to Ref. \cite{Z1}, one has the following well-known result
\begin{eqnarray}
\rho(\mu)&=&-\frac{Z_2}{{\cal{V}}}\mbox{Tr}\left\{G_R[\mu]\gamma_4
\right\}=-N_cN_f Z_2
\int\frac{d^4p}{(2\pi)^4}\mbox{tr}\left\{G_R[\mu](p)\gamma_4
\right\},\label{eq:def1}
\end{eqnarray}
where $G_R[\mu](p)$ is the renormalized quark propagator at finite
chemical potential and $N_{c}$, $N_f$ denote the number of colors and
flavors, respectively. Here, Tr denotes trace operation over color,
flavor, Dirac and coordinate indices, while tr denotes trace
operation over Dirac indices only.

Integrating both sides of the equation $\rho(\mu)=\partial\cal{
P}(\mu)/\partial\mu$ gives
\begin{equation}
{\cal{P}}(\mu)=\left.{\cal{P}}(\mu)\right|_{\mu=0}+
\int_{0}^{\mu}d\mu'\rho(\mu')=\left.{\cal{P}}(\mu)\right|_{\mu=0}
-N_cN_fZ_2\int_{0}^{\mu}d\mu'\int\frac{d^4p}{(2\pi)^4}
tr\left\{G_R[\mu'](p)\gamma_4\right\},\label{eq:def2}
\end{equation}
where $\left.{\cal{P}}(\mu)\right|_{\mu=0}$ is an integration
constant which represents the pressure density at $\mu=0$. Here one should note that
$\left.{\cal{P}}(\mu)\right|_{\mu=0}$ is a constant independent of
$\mu$ and the whole nontrivial $\mu$ dependence of ${\cal{P}}(\mu)$
is contained in the integration term which is totally determined by
the renormalized quark propagator $G_R[\mu](p)$. Formally Eq.
(\ref{eq:def2}) provides a model-independent formula for calculating
the pressure density at finite $\mu$. However, at present we still do not know how to calculate the quark
propagator at finite chemical potential from first principles of QCD. So, when one uses Eq. (\ref{eq:def2}) to calculate the pressure density, one has to resort to various QCD models or approximations.
In this paper we will use the HDL approximation.

\begin{figure}[ht]
\centering
\includegraphics[width=8cm]{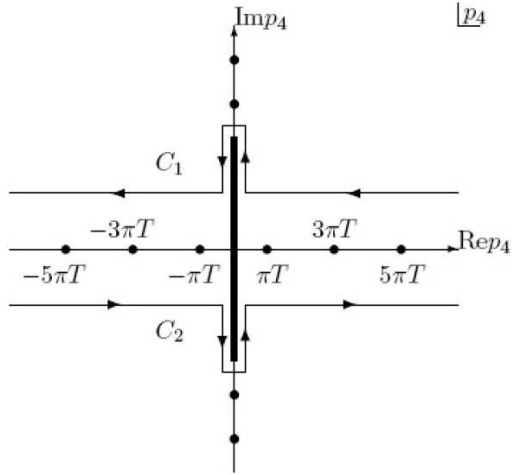}
\caption{The integral contour}\label{fig:qpAna}
\end{figure}

The quark propagator under HTL/HDL approximation can be written as \cite{Braaten}
\begin{eqnarray}
\label{eq:QPHTL1}
G_R(p)&=&-Z_2^{-1}\frac{1}{D_+(p)}\frac{\gamma_4+i\hat{\textbf{p}}\cdot
\vec{\gamma}}{2}-Z_2^{-1}\frac{1}{D_-(p)}\frac{\gamma_4-i\hat
{\textbf{p}}\cdot \vec{\gamma}}{2},
\end{eqnarray}
where $\hat{\textbf{p}}=\vec{p}/|\vec{p}|$ and $p_4=(2n+1)\pi T~ (n
\in {\bf Z})$ are the fermion Matsubara frequencies. The form of the
functions $D_\pm(p)$ is
\begin{eqnarray}
D_{\pm}(p)&=&-ip_4+\mu\pm |\vec{p}|+\frac{m_q^2}{|\vec{p}|}
\left[Q_0\left(\frac{ip_4-\mu}{|\vec{p}|}\right)\mp
Q_1\left(\frac{ip_4-\mu} {|\vec{p}|}\right)\right],
\end{eqnarray}
where $m_q\equiv g\sqrt{(T^2+\mu^2/\pi^2)/6}$ is the quark thermal
mass with $g$ being the strong coupling constant, $Q_0$ and $Q_1$
are Legendre functions of the second kind. Substituting Eq. (\ref{eq:QPHTL1}) into Eq. (\ref{eq:def1}) gives
\begin{eqnarray}
\rho(\mu)&=&2N_cN_f\int\frac{d^3\vec{p}}{(2\pi)^3}T\sum_n
\left[\frac{1}{D_+}+\frac{1}{D_-}\right]\label{eq:QDen2}.
\end{eqnarray}
According to a familiar method in thermal field theory, the frequency
sum in Eq. (\ref{eq:QDen2}) can be calculated by means of contour integral
\begin{eqnarray}
&&T\sum_n \left[\frac{1}{D_+}+\frac{1}{D_-}\right]+
\int_{-i|\vec{p}|-i\mu}^{i|\vec{p}|-i\mu}\frac{dp_4}{2\pi}
\mbox{Disc}\left[\frac{1}{D_+}+
\frac{1}{D_-}\right]\frac{1}{2}\tanh\frac{ip_4}{2T}\nonumber\\
&=&\int_{C_1\cup C_2}\frac{dp_4}{2\pi}\left[\frac{1}{D_+}+
\frac{1}{D_-}\right]\frac{1}{2}\tanh\frac{ip_4}{2T}.
\end{eqnarray}
The integral contour is shown in Fig. \ref{fig:qpAna}. Now let us
calculate the integral. Defining
\begin{eqnarray}
f(p_4)&:=&\left[\frac{1}{D_+}+
\frac{1}{D_-}\right]\frac{1}{2}\tanh\frac{ip_4}{2T}
\end{eqnarray}
and closing $C_1$ and $C_2$ with large half circles, one obtains the
following
\begin{eqnarray}
\int_{C_1\cup C_2}\frac{dp_4}{2\pi}\left[\frac{1}{D_+}+
\frac{1}{D_-}\right]\frac{1}{2}\tanh\frac{ip_4}{2T}
&=& \frac{1}{2\pi}(-2\pi i)\sum_j\mbox{Res}\{f(\zeta_j)\},
\label{eq:QDInte1}
\end{eqnarray}
where $\zeta_j$ are the poles of $f(p_4)$ located in the imaginary
axis. The poles of $1/D_+$ are
\begin{eqnarray}
&&\zeta_1=-i(\omega_++\mu)\,,\;\;\mbox{Res}\left\{\frac{1}{D_+(\zeta_1)}\right\}=iZ_+\\
&&\zeta_2=i(\omega_--\mu)\,,\;\;\;\;\;\mbox{Res}\left\{\frac{1}{D_+(\zeta_2)}\right\}=iZ_-
\end{eqnarray}
and the poles of $1/D_-$ are
\begin{eqnarray}
&&\zeta_3=i(\omega_+-\mu)\,,\;\;\;\;\;\mbox{Res}\left\{\frac{1}{D_-(\zeta_3)}\right\}=iZ_+\\
&&\zeta_4=-i(\omega_-+\mu)\,,\;\;\mbox{Res}\left\{\frac{1}{D_-(\zeta_4)}\right\}=iZ_-,
\end{eqnarray}
where $\omega_\pm(|\vec{p}|)(>|\vec{p}|)$ are the solutions of the
equations
\begin{eqnarray}
\frac{|\vec{p}|(\omega_+-|\vec{p}|)}{m_q^2}-1&=&\frac{1}
{2}\left(1-\frac{\omega_+}{|\vec{p}|}\right)\ln\frac
{\omega_++|\vec{p}|}{\omega_+-|\vec{p}|},\\
\frac{|\vec{p}|(\omega_-+|\vec{p}|)}{m_q^2}+1&=&\frac{1}
{2}\left(1+\frac{\omega_-}{|\vec{p}|}\right)\ln\frac
{\omega_-+|\vec{p}|}{\omega_--|\vec{p}|}
\end{eqnarray}
and $Z_\pm(|\vec{p}|)$ are
\begin{eqnarray}
Z_\pm(|\vec{p}|)&=&\frac{\omega_\pm-|\vec{p}|^2}{2m_q^2}.
\end{eqnarray}
From these results one can calculate the right hand side of Eq. (\ref{eq:QDInte1}):
\begin{eqnarray}
&&\int_{C_1\cup C_2}\frac{dp_4}{2\pi}\left[\frac{1}{D_+}+
\frac{1}{D_-}\right]\frac{1}{2}\tanh\frac{ip_4}{2T}=\frac{1}{2\pi}(-2\pi i)\sum_j\mbox{Res}\{f(\zeta_j)\}\\
&=&-\frac{i}{2}\left[iZ_+\tanh\frac{\omega_++\mu}{2T}+iZ_-
\tanh\frac{-(\omega_--\mu)}{2T}+iZ_+\tanh\frac{-(\omega_+-\mu)}{2T}
+iZ_-\tanh\frac{\omega_-+\mu}{2T}\right].\nonumber
\end{eqnarray}
The second term in the left hand side of Eq. (9) can be expressed as
\begin{eqnarray}
&&\int_{-i|\vec{p}|-i\mu}
^{i|\vec{p}|-i\mu}\frac{dp_4}{2\pi}\mbox{Disc}\left(\frac{1}{D_+}+
\frac{1}{D_-}\right)\frac{1}{2}\tanh\frac{ip_4}{2T}=\int_{-|\vec{p}|}^ {|\vec{p}|}
\frac{d\omega}{2\pi}[\rho_+(\omega)+\rho_-(\omega)]
\frac{1}{2}\tanh\frac{\omega-\mu}{2T},
\end{eqnarray}
where $\rho_\pm(\omega,|\vec{p}|)$ are the familiar spectral
functions of the quark propagator $(x\equiv\omega/|\vec{p}|)$
\cite{Braaten}:
\begin{eqnarray}
\rho_\pm(\omega,|\vec{p}|)&=&2\pi[Z_\pm(|\vec{p}|)\delta(\omega-
\omega_\pm(|\vec{p}|))+Z_\mp(|\vec{p}|)\delta(\omega+
\omega_\mp(|\vec{p}|))]\nonumber\\
&&+\frac{\pi m_q^2(1\mp
x)\theta(1-x^2)}{|\vec{p}|^3}\left\{\left[1\mp x+\frac{m_q^2}
{|\vec{p}|^2}\pm\frac{m_q^2}{2|\vec{p}|^2}(1\mp x)
\ln\frac{1+x}{1-x}\right]^2\right.\nonumber\\
&&\left.+\frac{\pi^2m_q^4}{4|\vec{p}|^4}(1\mp x)^2\right\}^{-1}.
\end{eqnarray}

Combining Eqs. (19), (20) and (9), one can obtain
\begin{eqnarray}
T\sum_n \left[\frac{1}{D_+}+\frac{1}{D_-}\right]&=&Z_+\bigg[n_F(\omega_+-\mu)-n_F(\omega_++\mu)\bigg]
+Z_-\bigg[n_F(\omega_-+\mu)-n_F(\omega_--\mu)\bigg]\nonumber\\
&&-\frac{1}{2}\int_{-|\vec{p}|}^ {|\vec{p}|}
\frac{d\omega}{2\pi}[\rho_+(\omega)+\rho_-(\omega)]
[1-2n_F(\omega-\mu)],\label{eq:SumR1}
\end{eqnarray}
where $n_F(\omega)$ is the Fermi distribution function
\begin{eqnarray}
n_F(\omega)&=&\frac{1}{\exp(\omega/T)+1}.
\end{eqnarray}
Therefore the quark number density at finite $\mu$ and $T$ is
\begin{eqnarray}
\rho(\mu,T)&=&2N_cN_f\int\frac{d^3\vec{p}}{(2\pi)^3}\left\{
Z_+\bigg[n_F(\omega_+-\mu)-n_F(\omega_++\mu)\bigg]\right.\nonumber\\
&&\left.+Z_-\bigg[n_F(\omega_-+\mu)-n_F(\omega_--\mu)\bigg]\right\}\nonumber\\
&&-N_cN_f\int\frac{d^3\vec{p}}{(2\pi)^3}\int_{-|\vec{p}|}^
{|\vec{p}|} \frac{d\omega}{2\pi}[\rho_+(\omega)+\rho_-(\omega)]
[1-2n_F(\omega-\mu)].\label{eq:QDen}
\end{eqnarray}
From Eq. (\ref{eq:QDen}) one can clearly see that the quark number density contains two terms, one is the
contribution from the quasi-particle poles and the other is the
contribution from the Landau damping. In the limit of $T\rightarrow\infty$,
$\rho$ tends to the free quark gas result ($Z_+\rightarrow1$, $Z_-\rightarrow0$
and the integral is dominated by $p\sim\omega\sim T$ region).

Before proceeding the calculation of EOS, let us turn to the
calculation of the QNS, which is defined as
\begin{eqnarray}
\chi(\mu,T)&\equiv&\frac{\partial\rho(\mu,T)}{\partial\mu}.\label{eq:QDen4}
\end{eqnarray}
As mentioned before, the QNS plays an important role in identifying
the critical end point in the QCD phase diagram. There are many
calculations of QNS under HTL approximation in the literature
\cite{Blaizot,Blaizot1,Chakraborty,Chakraborty1,HTL}. Generally, in order to calculate the QNS, one
has to calculate the quark propagator and the quark-meson vertex in
the vector meson channel at zero total momentum separately. In Ref.
\cite{HTL}, using the fact that $\partial/\partial\mu$
can be replaced by $\partial/\partial(-ip_4)$ and with the help of the
Ward-Takahashi identity, the authors calculate the QNS under HTL approximation only through the HTL quark propagator. Here a question arises: is the information of $\mu$ dependence lost by such a method in Ref. \cite{HTL}? Now we will answer this question in a straightforward way by calculating
the QNS directly from Eqs. (\ref{eq:QDen}) and (\ref{eq:QDen4}).
Because $Z_\pm$, $\rho_\pm$ and $\omega_\pm$ depend on $\mu$ only through $m_q$, one would obtain the following
\begin{eqnarray}
\chi(\mu,T)&=&\frac{\partial\rho(\mu,T,m_q)}{\partial\mu}+
\frac{\partial\rho(\mu,T,m_q)}{\partial m_q} \frac{\partial
m_q}{\partial\mu}.
\end{eqnarray}
Setting $\mu=0$ in the above equation gives
\begin{eqnarray}
\chi(\mu=0,T)&=&\left.\frac{\partial\rho(\mu,T,m_q)}{\partial\mu}
\right|_{\mu=0}+\left[\frac{\partial\rho(\mu,T,m_q)}{\partial m_q}
\frac{\partial m_q}{\partial\mu}\right]_{\mu=0}.\label{eq:QNS1}
\end{eqnarray}
One can easily find that the second term in Eq. (\ref{eq:QNS1}) is
zero ($(\partial m_q/\partial\mu)_{\mu=0}=0$) and the QNS at finite
$T$ and zero $\mu$ is
\begin{eqnarray}
\chi(\mu=0,T)&=&\frac{4N_cN_f}{T}\int
\frac{d^3\vec{p}}{(2\pi)^3}\left[\frac{\omega_+^2-\vec{p}^2}
{2m_q^2}n_F(\omega_+)(1-n_F(\omega_+)) +\frac{\omega_-^2-\vec{p}^2}
{2m_q^2}n_F(\omega_-)(1-n_F(\omega_-))\right]_{\mu=0}\nonumber\\
&&+\frac{2N_cN_f}{T}\int\frac{d^3\vec{p}}{(2\pi)^3}
\int_{-|\vec{p}|}^
{|\vec{p}|}\frac{d\omega}{2\pi}\bigg\{[\rho_+(\omega)+\rho_-(\omega)]
n_F(\omega)(1-n_F(\omega))\bigg\}_{\mu=0}.
\end{eqnarray}
This is the same as the result in Ref. \cite{HTL}, as one expected in advance. Here it should
be stressed that the method used to calculate the QNS in the present
paper is different from the one used in Ref. \cite{HTL} in which the Ward-Takahashi identity is adopted to avoid the differentiation over $\mu$. The result obtained here can be regarded as a self-check of the calculation in Ref. \cite{HTL}. As is pointed out in Ref. \cite{HTL}, the QNS contains the contributions from both the Landau damping and the quasi-particle poles. This is different from the result in Ref. \cite{Chakraborty} (see Ref. \cite{HTL} for more detail).

After the discussion of QNS let us continue the calculation of EOS.
The zero temperature result of $\rho$ reads
\begin{eqnarray}
\rho(\mu)&=&2N_cN_f\int\frac{d^3\vec{p}}{(2\pi)^3}\left[
\theta(\mu-\omega_+)Z_+-\theta(\mu-\omega_-)Z_-\right]\nonumber\\
&&+N_cN_f\int\frac{d^3\vec{p}}{(2\pi)^3}\int_{-|\vec{p}|}^
{|\vec{p}|} \frac{d\omega}{2\pi}[\rho_+(\omega)+\rho_-(\omega)]
[1-2\theta(\omega-\mu)].\label{eq:QDen3}
\end{eqnarray}
It should be pointed out that according to Eq. (4), in order to calculate the quark number
density one needs to know the quark propagator in the whole momentum range. However, the HTL/HDL approximation is only valid for external momentum much smaller than $T$ or $\mu$. Therefore, as in
Ref. \cite{HTL}, for a consistent calculation of the quark number
density under the HDL approximation one should introduce a momentum
cutoff $\Lambda_{HDL}$ below which the HDL quark propagator (6) is
applicable. In our calculation we choose $\Lambda_{HDL}=g\mu/\pi$.
We make this choice because $g\mu/\pi$ is an important energy scale
to identify the soft momentum in HDL approximation (for HTL this
would be $gT$, see Ref. \cite{HTL}). If we limit the range of
integration of the contribution of Landau damping in Eq.
(\ref{eq:QDen3}) to the region $|{\vec p}|\leq g\mu/\pi$, we will
get the numerical result of the quark number density shown in Fig. 2
($N_c=3,N_f=2$). From Fig. 2 it can be seen that the quark number density
under HDL approximation is slightly smaller than that of free quark
gas ($\rho_{free}(\mu)=N_cN_f\mu^3/3\pi^2$), which indicates that the attractive interaction dominate in the high
density case. Here we want to stress that the result in Fig. 2 has minor sensitivity
to the variation of the cut-off $g\mu/\pi$. For
example, if one sets the cut-off to be $2g\mu/\pi$, one finds that the
change of $\rho$ is about $10\%$. This is because the quark number
density (and susceptibility) is determined essentially by the quasi-particle contribution
within the given approximations.
\begin{figure}[ht]
\includegraphics[width=9cm]{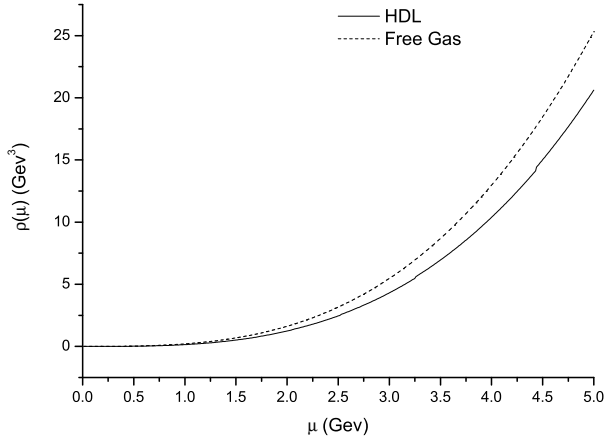}\\
\caption{The quark number density under HDL approximation}
\end{figure}

Now we can calculate the EOS under HDL approximation (here we neglect
the constant term ${\cal P}(\mu)|_{\mu=0}$ because when applying our
EOS to the study of neutron star, owing to the boundary condition
imposed on the surface of neutron star, the constant term ${\cal
P}(\mu)|_{\mu=0}$ does not contribute to the mass-radius relation)
\begin{eqnarray}
\mathcal{P}(\mu)&=&2N_cN_f\int_0^\mu d\mu^\prime
\int\frac{d^3\vec{p}}{(2\pi)^3}\left[
\theta(\mu^\prime-\omega_+)Z_+-\theta(\mu
^\prime-\omega_-)Z_-\right]\nonumber\\
&&+N_cN_f\int_0^\mu
d\mu^\prime\int\frac{d^3\vec{p}}{(2\pi)^3}\int_{-|\vec{p}|}^
{|\vec{p}|}\frac{d\omega}{2\pi}[\rho_+(\omega)+\rho_-(\omega)]
[1-2\theta(\omega-\mu^\prime)].
\end{eqnarray}
The numerical results are shown in Fig. 3 with $P_{free}=N_cN_f\mu^4/(12\pi^2)$ being the pressure density of the free quark gas. It can be seen that the pressure density under the HDL approximation is smaller than that of the free quark gas.

So far we have derived an expression of the nontrivial $\mu$ dependence of the pressure density in the framework of the HDL approximation. Now let us give a discussion of the features of our approach. 
The study of EOS of quark matter is a long-standing subject of QCD study. 
Just as was mentioned in the introduction, although lattice QCD has achieved some success in treating small
$\mu$ physics, they are far from being complete in the case of large chemical potential problem. 
So one expects that continuum model studies should be complementary to the lattice simulations in the
study of QCD thermodynamics at finite $\mu$.
When the chemical potential is large enough, one may naively expects that perturbative calculation is feasible because of asymptotic freedom. However, as was pointed out in Ref. \cite{Blaizot2}, the perturbative calculation of partition function of QCD at finite $T$ or $\mu$ encounters the convergence problem and one has to resort to resummation techniques. From the point of view of functional integral approach, the calculation of partition function of QCD amounts to the calculation of all possible vacuum bubble diagrams. This is a hard work. In order to avoid the problem of calculating the complicated vacuum bubble diagrams, in our work we adopt the approach proposed in Ref. \cite{Z1}, in which the nontrivial $\mu$ dependence of the pressure density is totally determined by the full quark propagator at finite $\mu$ (the quark propagator is the simplest Green function of QCD). If the input dressed quark propagator at finite $\mu$ is exact, the pressure density calculated with the aid of it will also be exact. At present it is not possible to determine reliably the dressed quark propagator at finite 
$\mu$ from first principles of QCD. So one has to resort to various models inspired by QCD. It is generally believed that the HTL/HDL approximations are good approximations of QCD when $T$ or $\mu$ is large enough. So in the present work we employ the HDL approximation to calculate the EOS of quark matter. Compared with other approaches of calculating the EOS of quark matter in the literature, our approach is simpler and has a better QCD foundation.

The behavior of the pressure density under HDL approximation shown
in Fig. 3 is qualitatively differs from that of the naive perturbation
theory. For example, in Ref. \cite{FPS} Fraga, Pisarski and
Schaffner-Bielich (FPS) proposed the following cold, perturbative EOS of QCD
\begin{equation}
{\cal P}_{FPS}(\mu)=\frac{N_f\mu^4}{4\pi^2}\Bigg\{1-2\Big(
\frac{\alpha_s}{\pi}\Big)-\Bigg[G+N_f \ln
\frac{\alpha_s}{\pi}+\Big(11-\frac{2}{3}N_f \Big)\ln\frac{{\bar
\Lambda}}{\mu}\Bigg]\Big(\frac{\alpha_s}{\pi}\Big)^2\Bigg\},
\end{equation}
where $G=G_0-0.536N_f+N_f \ln N_f$, $G_0=10.374 \pm 0.13$ and ${\bar
\Lambda}$ is the renormalization subtraction point. The scale
dependence of the strong coupling constant $\alpha_s({\bar
\Lambda})$ is taken as
\[
\alpha_s({\bar \Lambda})=\frac{4\pi}{\beta_0 u}\Bigg[
1-\frac{2\beta_1}{\beta_0^2}\frac{\ln(u)}{u}+\frac{4\beta_1^2}
{\beta_0^4
u^2}\Bigg(\Big(\ln(u)-\frac{1}{2}\Big)^2+\frac{\beta_2\beta_0}
{8\beta_1^2}-\frac{5}{4}\Bigg) \Bigg],
\]
where $u=\ln({\bar \Lambda}^2/\Lambda_{\overline{MS}}^2)$,
$\beta_0=11-2N_f/3$, $\beta_1=51-19N_f/3$, and
$\beta_2=2857-5033N_f/9+325N_f^2/27$. For $N_f=3$,
$\Lambda_{\overline{MS}}=365~\mathrm{MeV}$. The only freedom in the
model of Ref. \cite{FPS} is the choice of the ratio ${\bar
\Lambda}/\mu$, which is taken to be 2 in that reference. The
comparison between the HDL result and FPS result is shown in Fig. 3.
From Fig. 3 it can be seen that when $\mu$ is smaller than about $
1$ GeV, the HDL pressure density is smaller than the perturbative
result to the $\alpha_s$ order, but is larger than that to the
$\alpha_s^2$ order. When $\mu$ is larger than about $1$ GeV, the HDL
pressure density is smaller than the FPS one, and an important result
is that when $\mu$ tends to infinity, the HDL pressure density tends
to the free quark gas result much more slowly than does the FPS one. As a
comparison, the EOS obtained under Dyson-Schwinger equations (DSEs)
approach \cite{Z1} with different parameters are also shown in Fig. 3.
\begin{figure}[ht]
\includegraphics[width=9cm]{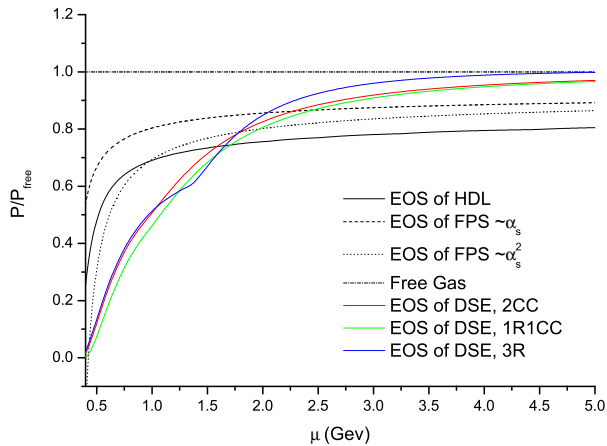}\\
\caption{The different EOS}
\end{figure}

It should be noted that in drawing the EOSs shown in Fig. 3, we have neglected the term $\mathcal{P}(\mu)|_{\mu=0}$. This does not mean that the term $\mathcal{P}(\mu)|_{\mu=0}$ is unimportant. In fact, it is an important quantity. It enters the energy density, which is relevant for integrating the Tolman-Oppenheimer-Volkoff equations. At present it is not possible to calculate reliably $\mathcal{P}(\mu)|_{\mu=0}$  from first principles of QCD. One can only calculate it using various models inspired by QCD, for example, using the CJT effective action (see, for instance, Ref. \cite{Zong2}). It is obvious that the existing calculations of $\mathcal{P}(\mu)|_{\mu=0}$ in the literatures are all model dependent. In our work, we have not considered the term $\mathcal{P}(\mu)|_{\mu=0}$. The main reason is that this term does not affect the relative relation of different EOSs displayed in Fig. 3, since all of the curves shown in Fig. 3 do not contain this term. If one tries to add the term $\mathcal{P}(\mu)|_{\mu=0}$ in the EOS and then make comparison between different EOSs, one has to add it in all different EOSs. Therefore the main conclusion of the present paper does not change.

Now it is time to discuss the range of applicability of the EOS presented in this paper. For the low temperature and high chemical potential regime, our knowledge is quite limited. Experimentally, it is impossible to achieve such a condition in laboratories on earth. A natural laboratory holding such cold, highly compressed matter is a quark star. So the only obvious relevance of the presented EOS is for quark stars. It is well known that in astrophysics the study of quark star depends crucially on the assumed EOS \cite{N1,N2}. Therefore it is interesting to apply the present EOS to the study of quark star. In our future work \cite{Li} we shall consider the structure of the quark star for the three flavor situation, where we shall study the mass-radius relation, mass-central-density relation, and distribution of inner mass and chemical potential for this kind of quark star. In addition, we shall compare the results calculated from the present EOS with those calculated from other EOSs in Fig. 3. 

To summarize, in this paper we calculate the EOS of QCD at zero
temperature and finite quark chemical potential under the HDL
approximation. It is found that when $\mu$ tends to infinity the HDL
pressure density tends to the free quark gas result much more slowly
than that of the cold, perturbative EOS of QCD of Fraga, Pisarski
and Schaffner-Bielich. We also give the expression of QNS at finite
$T$ and $\mu$. It is found that when $\mu\rightarrow0$, the result of QNS is
the same as the one obtained in Ref. \cite{HTL}.

\begin{acknowledgements}
We thank J.-P. Blaizot for discussions on HTL and HDL approximation. This work is supported in part by the National Natural Science Foundation of China (under Grant Nos. 10775069 and 10935001) and the Research Fund for the Doctoral Program of Higher Education (Grant Nos. 20060284020 and 200802840009).
\end{acknowledgements}

\end{document}